# Development of Rebunching Cavities at IAP

C. P. Welsch, K.-U. Kühnel, A. Schempp
Institut für Angewandte Physik
*Johann Wolfgang Goethe Universität, Frankfurt am Main, Germany*

*Abstract*

A focus of work at IAP has been the development and optimization of spiral loaded cavities since the 1970s [A. Schempp et al, NIM **135**, 409 (1976)]. These cavities feature a high efficiency, a compact design and a big variety of possible fields of application.

They find use both as bunchers and post accelerators to vary the final energy of the beam. In comparison to other available designs, the advantage of these structures lies in their small size. Furthermore they can easily be tuned to the required resonance frequency by varying the length of the spiral. Due to the small size of the cavities the required budget can also be kept low.

Here, two slightly different types of spiral loaded cavities, which were built for the REX-ISOLDE project at CERN and the intensity upgrade program at GSI are being discussed.

## I. COMMON FEATURES

### A. General Remarks

Large Accelerator facilities very often need to be highly flexible in terms of final energy and the particles to be accelerated. With long linear accelerators – like Alvarez or Widerøe structures – alone, the necessary energy variations cannot be attained. Therefore, single resonators are very often used. They allow changing the energy of the particles as well as the velocity distribution of the beam.

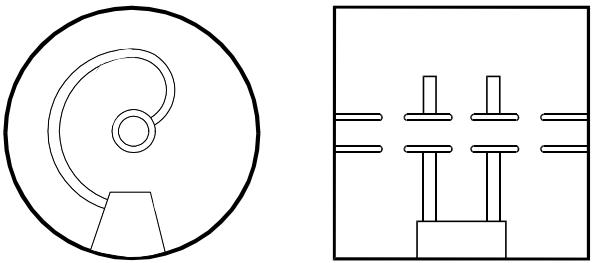

Fig. 1: Schematic Drawing of spiral loaded cavity

In a cylindrical external tank one or more spiral arms are placed. At their open end each spiral carries a drift tube running along the symmetry axis of the cavity and thus forming several accelerating gaps.

Different designs of the spiral arms are possible, depending on the required resonance frequency. The most common structure consists of two semi-circles with radii r and r/2. This design allows easy manufacturing as well as simple adjustment inside the tank – the drifttubes are automatically situated in the middle of the cavity. Somehow more complicated are the "archimedic Type" structures. In this case spirals with constant pitch are used allowing resonance frequencies as low as 27 MHz.

### B. Characteristic Parameters

The transittime factor $T$ indicates the flexibility of an accelerator as the quotient of the effectively experienced voltage and the total gap voltage. The $R_p$-value is defined as the effective cavity voltage squared over the power $P$ put into the cavity.

$$T = \frac{U_{eff,Gap}}{U_{tot}} \qquad R_p = \frac{U^2_{eff,Cav.}}{P}$$

With $n$ as the number of gaps, the $R_p$-value is proportional to $n^2$ in case of the single-spiral structure and proportional to $n$ for a splitring structure.

The transittime factor is the product of a velocity dependent and a geometry based term [1]

$$T_G = \frac{\beta \lambda}{\pi(d + 0,85 \cdot r)} \cdot sin\left(\frac{\pi(0,85 \cdot r)}{\beta \lambda}\right) \cdot \frac{1}{I_0\left(\frac{2\pi a}{\beta \lambda}\right)}$$

$$T_V = \frac{2 \cdot \beta}{n \cdot \pi \cdot (\beta - \beta_0)} \cdot sin\left(\frac{n \cdot \pi \cdot (\beta - \beta_0)}{2 \cdot \beta}\right)$$

where  $d$ = width of acceleration gap
 $r$ = radius of curvature of drifttubes ($r_{out}$-$r_{in}$)
 $a$ = radius of aperture
 $\beta_0 = v_0/c;\ \beta = v/c$
 $v_0/v$ = reference / real velocity of particles

Fig. 2 shows a plot of the velocity dependent term in case of the 36 MHz 11.4 MeV/u rebuncher built for GSI.

The wide spectrum of possible applications is clearly shown in this plot. While many-gap structures are limited to a narrow energy interval, a 2- or 3-gap accelerator can handle even particles with a large deviation from the design energy. This flexibility is characteristic for all of the here presented structures.

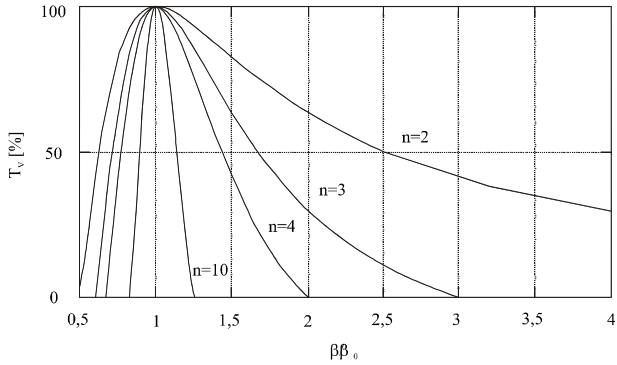

Figure 2: velocity depend term of the transittime factor

The resonance frequency of the structures can also be determined from the geometrical dimensions alone [2].
During the last years analytical solutions for describing these cavities have been continuously developed and allow the calculation of not only the transittime factor, but also the quality factor Q, the shunt impedance and resonance frequency with high accuracy.
Without going into the details, it should be mentioned that it is possible to describe a spiral loaded cavity as a transission line loaded with capacitors or as a set of capacitors and inductors. Only the geometric dimensions of cavity and the inner structure are necessary for these calculations and are therefore applicable to all different kinds of structures. Fig. 3 shows the calculated capacity $C$ at each *mm* of the 36 MHz 11.4 MeV/u spiral in the transfer channel at GSI. Included is not only the capacity between the spiral and the surrounding tank, but also between the two spirals and interaction between the windings of each arm.

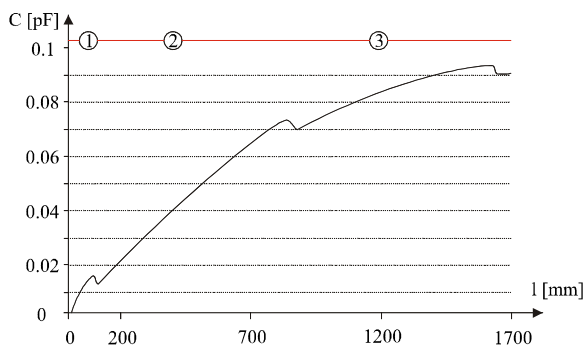

Figure 3: capacity $C$ between spiral arm and tank

In addition to these analytical calculations, all of the above parameters are determined with the MAFIA program. Using the obtained field distributions, the accuracy of the above results can be proven and optimisation of the structures can be done.

## II. THE INTENSITY UPGRADE PROGRAM AT GSI

It is the goal of the intensity upgrade program at GSI to fill the SIS up to its space charge limit. This requires a beam current of up to 15 emA ($^{238}U^{4+}$), which could not be delivered by the earlier system. Therefore the old wideroe structure was replaced by an RFQ and two IH-structures, especially designed to accelerate high intensity, low charge state beams.
Special attention had to be paid in the two stripping sections, where high space charge effects are present and focussing elements in both transverse and longitudinal direction are necessary.

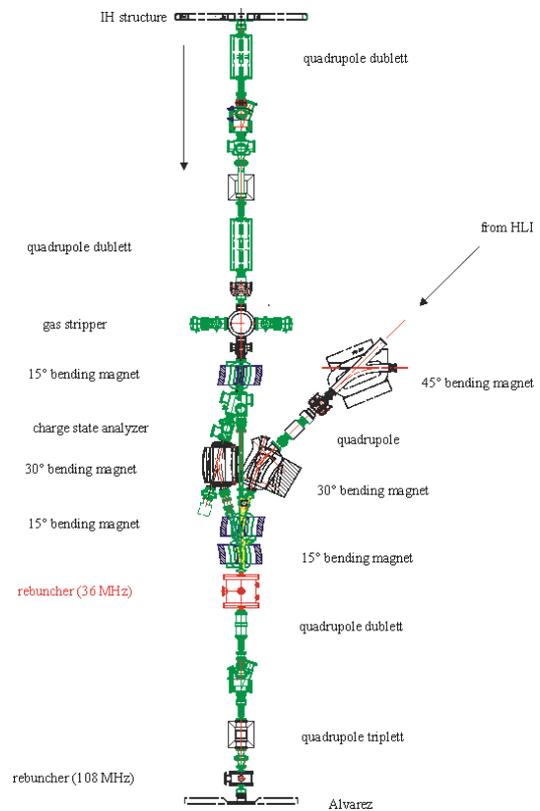

Figure 4: part of the UNILAC

Even though an 108 MHz Rebuncher was already present in the section in front of the Alvarez-linac, it was necessary to counteract the longitudinal defocusing effects.
The resonance frequency was chosen to be 36 MHz, like the frequency in the HSI, providing a larger region of linear forces.

This low frequency and the given length of the structure with only *l=60 cm* required a compact structure. At high currents a total voltage of 875 kV was necessary to turn the emittance ellipse into an upright position. In order to avoid sparks, two spirals with constant pitch ("archimedic type") were chosen, shown in Fig 5.

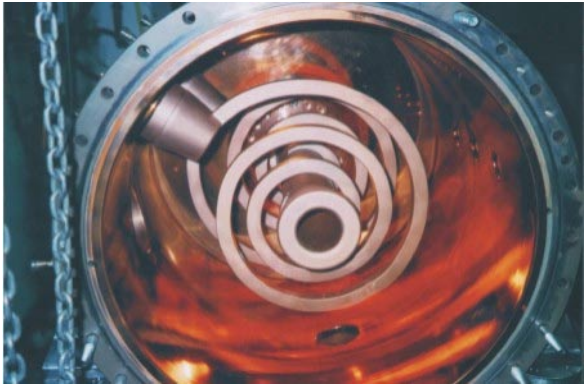

Figure 5: View inside 36 MHz 1.4 MeV/u rebuncher

The radius of the tank is only $r=25\ cm$, the aperture $d_{ap}=5.5\ cm$. Experiments showed that this structure was not stable against external mechanical excitations. With a length of $210\ cm$ each arm was not inflexible enough to resist forced oscillation. Thus two additional copper tubes were fixed on each side of the two spirals.

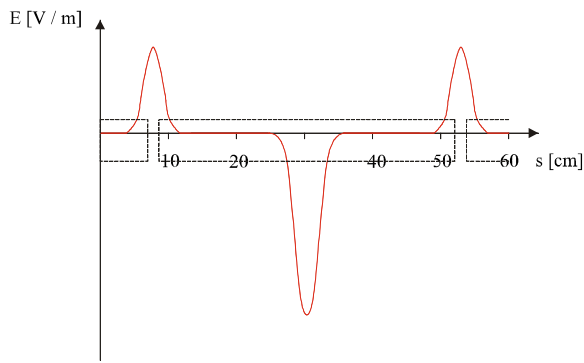

Figure 6: distribution of electric field along axis

A number of simulations was done to optimise the particle motion through the buncher [2,3]; always taking into account the various kinds of different ions that are going to be used at GSI.

Fig 6 shows the distribution of the electric field along the beam axis. Since all gaps have the same width, the field in the middle gap shows double the amplitude. With a quality factor of $Q_0=2200$ and $R_p=9\ M\Omega$ the input power needs to be $p=85\ kW$.

Another rebuncher had to be installed between the UNILAC and the SIS. In this transfer channel the ions pass through a carbon foil where they get many of their electrons stripped. Strong space charge effects are the result of this process leading to a longitudinal debunching. In order to match the beam into the SIS, a second rebuncher counteracts this divergence.

Operated at the same resonance frequency $f_{res}=36\ MHz$, the "archimedic type" structure was chosen again. Since the design energy of $E=11.4\ MeV/u$ was a lot higher in this case, the total voltage needed to be only $U=650 kV$ which could be fulfilled with a single spiral.

Each of the two gaps has a width of 2.2 cm and the diameter of the drifttubes is $d_{ap}=6\ cm$. The $l=170\ cm$ long spiral has an internal water cooling to secure stable conditions during operations.

The quality factor is a little higher in this case $Q_0=3600$ and with a measured value of $R_p=4.5\ M\Omega$, an input power of $p=94\ kW$ is needed.

Both resonators are already integrated into the GSI beamline and are successfully operated.

## III. REBUNCHER FOR REX-ISOLDE

REX-ISOLDE (Radioactive Beam Experiment at ISOLDE) is a new project at the online mass separator ISOLDE / CERN [4].

The structure of exotic and neutron rich nuclei, especially Na, Mg, K and Ca, will be studied. The nuclei have to reach energies in the order of the coulomb barrier.

By measuring the arising γ-rays both, static and dynamic properties of these hardly known nuclei can be examined.

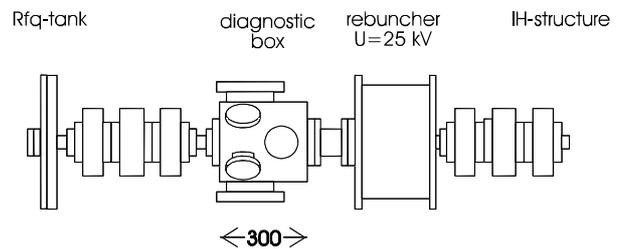

Figure 7: Part of the REX-ISOLDE beamline

In order to accelerate the radioactive beam to the necessary energy, a combination of an ion source, penning trap, charge state breeder, mass separator, an Rfq, an IH-structure and a 7-gap resonator is used [5]. Finally the radioactive beam hits the target with a final energy from 0.8 MeV/u to 2.2 MeV/u.

Part of the whole beamline is shown in Fig. 7. Between RFQ and IH a longitudinal match is necessary. Beam diagnostic is done and the beam is matched to the acceptance of the IH structure. Limiting parameter was the total length of only 1.6 m from the exit of the Rfq to the IH. Two quadrupole triplets, the diagnostic box and the rebuncher had to fitted in. Thus the maximum length of the rebuncher tank must not exceed 30 cm.

The necessary transformation of the phase space ellipse required a total voltage of 25 kV. Combining flexibility with high efficiency, a 3-gap splitring structure was chosen.

Care has to be taken to separate 0- and π-mode from one another. Calculation and Measurements showed that the unwanted 0-mode lies at $f_0=159$ MHz and therefore far enough from the operating frequency of $f_{res}=101.28$ MHz. Furthermore, the electric field needs to be kept in reasonable dimensions in order to avoid sparking. The result of the calculations is shown in Fig. 8.

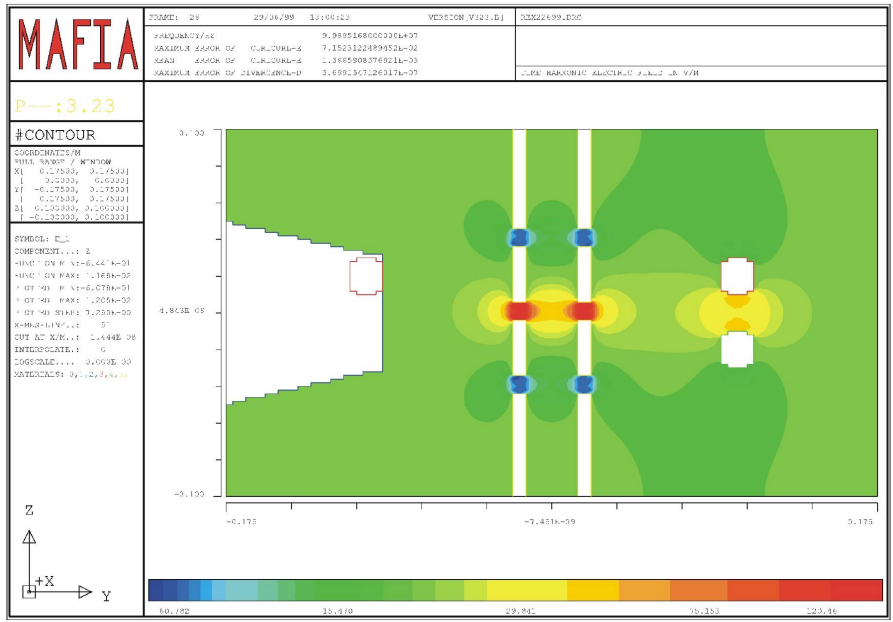

Figure 8: calculated field distribution

The inner diameter of the vacuum tank is 35 cm - incorporating two spirals with a length of only $l=475$ mm each. The beam aperture is $d_{ap}=6cm$ as in all other elements of the matching section.

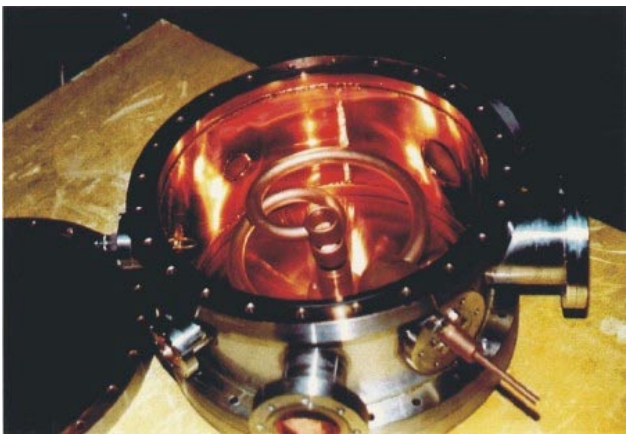

Figure 9: View into open splitring cavity

This structure was also completed and is today operated successfully in Munich and will be integrated in the REX-ISOLDE beamline at CERN soon.

## IV. CONCLUSION AND OUTLOOK

During the last three decades, the efficiency and flexibility of rebunching cavities has been more and more developed. The compact size of these cavities clearly separates them from other possible designs.

Easy manufacturing, low input power and stability during operation are the features of these cavities that can be used as rebunchers as well as post accelerators.
Different internal structures are possible and depend on the resonance frequency, energy region and the possible size.
Future research will optimise present designs with respect to required power input, possible voltages and size.